 \documentclass[final,5p,times,twocolumn]{elsarticle}

\biboptions{sort&compress}

  \usepackage{graphicx}
  \usepackage{epstopdf}
  \usepackage{verbatim}

\usepackage{amssymb}
\usepackage{color}

\journal{Physics Letters A}

\begin{document}

\begin{frontmatter}



\title{Instability and noise-induced thermalization of Fermi-Pasta-Ulam recurrence in the nonlinear Schr\"odinger equation}


 \author[Brescia]{Stefan Wabnitz}
 \ead{stefan.wabnitz@unibs.it}
 \author[Montreal]{Benjamin Wetzel}

 \address[Brescia]{Dipartimento di Ingegneria dell'Informazione,  Universit\`a degli Studi di Brescia, via Branze 38, 25123, Brescia, Italy}
 \address[Montreal]{INRS-EMT, 1650 Blvd. Lionel-Boulet, Varennes, Qu\'ebec J3X 1S2, Canada}

\begin{abstract}
We investigate the spontaneous growth of noise that accompanies the nonlinear evolution of seeded modulation instability into  Fermi-Pasta-Ulam recurrence. Results from the Floquet linear stability analysis of periodic solutions of the three-wave truncation are compared with full numerical solutions of the nonlinear Schr\"odinger equation. The predicted initial stage of noise growth is in a good agreement with simulations, and is expected to provide further insight into the subsequent dynamics of the field evolution after recurrence breakup.
\end{abstract}


\end{frontmatter}



\section{Introduction}
\label{sec:intro}
Modulation instability (MI) in the evolution of constant amplitude background waves as described by the self-focusing nonlinear Schr\"odinger equation (NLSE) is a well-known phenomenon with dramatic impact in a variety of physical settings ranging from nonlinear optics, hydrodynamics, plasma physics and Bose-Einstein condensation \cite{bespa,benjamin}. Among others, MI is at the origin of the generation of optical solitons, supercontinuum \cite{dudley06, dudley09}, and associated extreme events (the so-called optical rogue waves) \cite{solli07}. Whenever MI is induced by a weak seed signal \cite{tai}, the initial stage of exponential sideband amplification is followed by a nonlinear stage characterized by the so-called Fermi-Pasta-Ulam (FPU) recurrence phenomenon \cite{fermi,zk}. In fact, the NLSE model predicts a fully reversible or spatially periodic power exchange among the pump, the initial modulation sidebands and all their harmonics. This process can be described in terms of exact solutions of the NLSE \cite{akhmediev86,akhmediev88a,akhmediev88b,ablo}, and it has been experimentally observed in deep water waves \cite{lake, chab11}, nonlinear optical fibers \cite{sima01,sima02,tang,hamma,kibler10, kibler12}, nematic liquid crystals \cite{beeckman}, magnetic film strip-based active feedback rings \cite{wu}, and bimodal electrical transmission lines \cite{farota}. Useful physical insight into the qualitative dynamics of FPU recurrence, such as for example the presence of separatrix solutions, or the  dependence of the FPU recurrence period upon the input relative phase between the pump and the initial modulation sidebands, may be obtained by using a three-wave truncation which leads to a fully integrable, one-dimensional equivalent particle model \cite{infeld81,capp91,trillo91}.
 
Although the nonlinear stage of MI is characterized by the FPU recurrence phenomenon, the development of optical supercontinuum (SC) is associated with an irreversible evolution towards a thermalization state, i.e., a nearly equal distribution of spectral energy among all frequency components \cite{dudley06, dudley09}. Indeed, it has been recently predicted and experimentally observed that third-order dispersion induced losses into Cherenkov radiation may lead to irreversible energy dissipation of the pump field, which ultimately breaks the FPU recurrence \cite{soto12,musso14}. As a matter of fact, experiments carried out so far in nonlinear fiber optics have not been able to demonstrate FPU recurrence beyond a single spatial period \cite{sima01,sima02,tang,hamma, kibler10}. In addition, recent studies regarding noise-induced MI have highlighted the complex dynamics associated with the onset stage of noise amplification \cite{solli12, wetz12} and their intriguing links with the mechanisms underlying rogue waves formation in optical fiber \cite{dudley09, solli08}. Among others, it was shown that seeding the initial stage of SC generation with a weak modulation could lead to the stabilization of the SC output \cite{solli08, dud08, nguyen13}, and thus reduce the impact of noise on the complex dynamics associated with SC and rogue wave generation.

In this Letter we show that, even in the purely conservative case, there is a fundamental instability mechanism which breaks the FPU recurrence and leads to the irreversible evolution into statistically stationary spectra.
For this purpose, we carry out a linear stability analysis of FPU recurrence. We do that by analyzing the evolution of small perturbations around the exact solutions of the three wave truncation involving the pump and its initial sidebands. It is worth mentioning that the stability of periodic nonlinear mode coupling was studied before in the context of polarization MI in birefringent fibers \cite{trillo97a}, parametric mixing \cite{trillo97b} or second harmonic generation in quadratic materials \cite{fuerst97}, and, more recently, in the closely related problem of dual-frequency pumped four-wave mixing (FWM) in optical fibers \cite{arma11,fatome13,arma14}. By comparing the predictions of the linearized stability analysis of a simple three-mode truncation with the full numerical solutions of the NLSE, we show that a good quantitative agreement may be obtained between the two approaches. This allows us to show that the intrinsic MI process underlying the initial FPU recurrence phenomenon leads to its eventual break-up and spectral thermalization in the presence of input noise. The physical mechanism behind the FPU recurrence break-up may thus be simply understood in terms of the competing growth of the spontaneous (or noise-activated) MI of the periodically evolving pump and FWM sidebands.       

\section{Analytical stability theory}
\label{sec:theo}

The propagation of an optical pulse with envelope $u$ in a weakly dispersive and nonlinear dielectric (e.g., an optical fiber) can be described using the NLSE
\begin{equation}
	\nonumber i\frac{\partial u}{\partial z}+\frac{1}{2}\frac{\partial^2 u}{\partial t^2}+|u|^2u=0.\label{nls1}
\end{equation}
\noindent Here $z$ and $t$ denote dimensionless distance and retarded time, respectively. We restrict our attention to the anomalous group-velocity dispersion (GVD) regime. In order to analyze the dynamics of induced MI, we consider the input conditions given by Eq.(\ref{input}) 
\begin{equation}
\nonumber u(z=0,t)=1+\epsilon \exp\left\{i\phi^0/2\right\} \cos(\Omega t),\label{input}
\end{equation}
\noindent which corresponds to a perturbed continuous wave (CW) solution whenever $\epsilon \ll 1$. As well known, the linear stability analysis of Eqs.(\ref{nls1})-(\ref{input}) predicts that MI occurs for $0<\Omega<2$, with a gain peak found for $\Omega=\sqrt{2}$. 

Let us study the nonlinear evolution of MI in terms of the simple three-wave truncation \cite{capp91,trillo91}
\begin{equation}
\nonumber  u(z,t)=A_0(z,t)+A_{-1}(z,t)\exp(i\Omega t)+A_{+1}(z,t)\exp(-i\Omega t),\label{ansatz}
\end{equation}
\noindent where $A_0$, $A_{-1}$ and $A_{+1}$ denote the amplitudes of the pump, Stokes and anti-Stokes waves, respectively. By inserting the ansatz (\ref{ansatz}) in (\ref{nls1}), one obtains:
\begin{eqnarray}
 i\frac{\partial A_0}{\partial z}+\frac{1}{2}\frac{\partial^2 A_0}{\partial t^2} +\nonumber\\
\left(|A_0|^2 + 2|A_{-1}|^2 +  2|A_{+1}|^2\right) A_0 + 2A_{-1}A_{+1}A_{0}^* = 0 \nonumber\\
i\frac{\partial A_{-1}}{\partial z} -\frac{\Omega^2}{2}A_{-1} + i \Omega\frac{\partial A_{-1}}{\partial t}+\frac{1}{2}\frac{\partial^2 A_{-1}}{\partial t^2} +\nonumber\\
\left(2|A_0|^2 + |A_{-1}|^2+2|A_{+1}|^2\right) A_{-1} + A_{+1}^* A_{0}^2 =  0 \nonumber\\
i\frac{\partial A_{+1}}{\partial z} -\frac{\Omega^2}{2}A_{+1} - i \Omega\frac{\partial A_{+1}}{\partial t}+\frac{1}{2}\frac{\partial^2   A_{+1}}{\partial t^2} +\nonumber\\
\left(2|A_0|^2 + 2|A_{-1}|^2 +|A_{+1}|^2\right) A_{+1} + A_{-1}^* A_{0}^2 =  0 
\label{twe} 
\end{eqnarray}

\noindent The stationary solution $A_j(z,t)=\bar{A}_j(z)$ of Eq.(\ref{twe}) is exactly integrable \cite{capp91}. Let us set $\eta(z)=\left|\bar{A}_0(z)\right|^2/P_0$ and $\phi=\phi_{-1}+\phi_{+1}-2\phi_0$, with $\bar{A}_j(z)=\left|\bar{A}_j(z)
\right|\exp\{i\phi_j(z)\}$, where the conserved total power $P_0=\left|\bar{A}_0\right|^2+\left|\bar{A}_{-1}\right|^2+\left|\bar{A}_{+1}\right|^2$. Supposing for simplicity initial equal amplitude sidebands $\bar{A}_{-1}(z=0)=\bar{A}_{+1}(z=0)$, one obtains the 1-dimensional Hamiltonian system which describes the spatial evolution of the coordinates $(\eta,\phi)$ in a phase plane
\begin{equation}
\nonumber  \frac{d\eta}{dZ}=\frac{dH}{d\phi},\;\frac{d\phi}{dZ}=-\frac{dH}{d\eta},\label{onedim}
\end{equation}
where $Z=P_0z$, and the conserved Hamiltonian $H$ reads as
\begin{equation}
\nonumber  H=2\eta\left(1-\eta\right)\cos(\phi)-\left(\kappa-1\right)\eta-\frac{3}{2}\eta^2,\label{ham}
\end{equation}
where $\kappa=-\Omega^2/P_0$. The solutions of Eqs.(\ref{onedim})-(\ref{ham}) can be expressed in terms of Jacobian elliptic or hyperbolic functions \cite{capp91}. Even though higher-order sideband growth is neglected and, accordingly, pump depletion is underestimated when following the three-wave truncation approach, it has been shown that Eqs.(\ref{onedim})-(\ref{ham}) provide a relatively good model for describing the induced MI process in the NLSE (\ref{nls1}) as long as $1\leq\Omega\leq 2$ \cite{infeld81,trillo91}. In this regime, the dynamics and the periodic evolution of all harmonics of the input modulation are driven by the three-wave truncation. Whereas for $0\leq\Omega < 1$, some higher-order sidebands (e.g., the first-harmonic of the input modulation $\pm2\Omega$ for $0.5\leq\Omega < 1$) may also experience exponential growth with distance whenever they fall within the MI bandwidth \cite{swab10}. In this case the field evolution may be obtained as a nonlinear superposition of all linearly unstable modes, which leads to the emergence of multiple spatial periods \cite{akhmediev85,erki11}.    

\begin{figure}[ht]
\centering
\includegraphics[width=8.5cm]{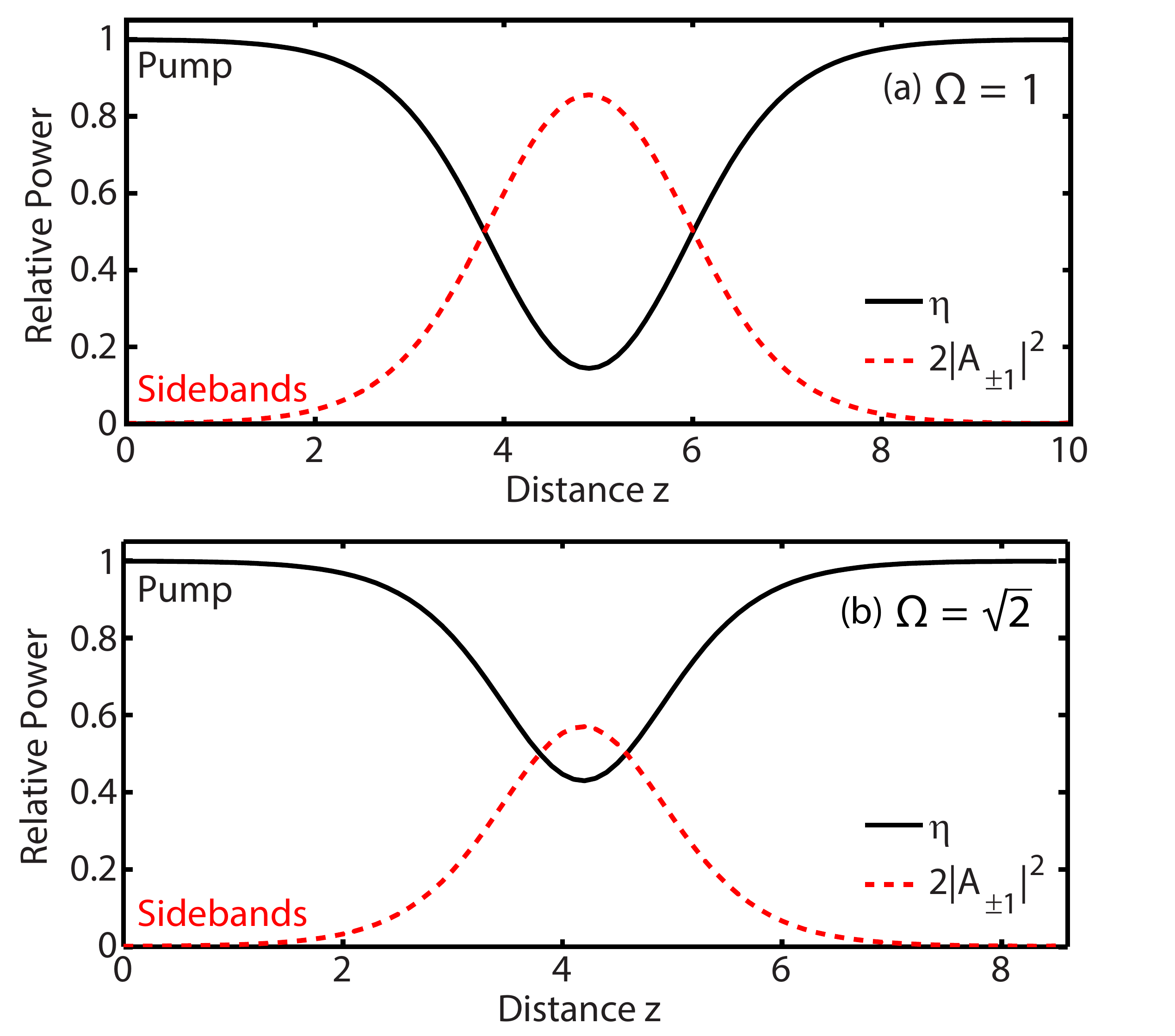}
\caption{Evolution of fractional power in the pump (solid black curves) and sidebands (dashed red curves) from 3-wave model for: (a) $\Omega=1$; (b) $\Omega=\sqrt{2}$.}
\label{power}
\end{figure}

Fig.\ref{power} shows examples of periodic solutions of Eqs.(\ref{onedim})-(\ref{ham}), which can be expressed in terms of an elliptic integral \cite{capp91}. Here we only display the evolution over one period, say, $z=\Lambda$. We have set $\eta\,(Z=0)=0.9988$ and $\phi\,(Z=0)=0$, which is equivalent to setting $\epsilon=0.05$ and $\phi^0=0$ in Eq.(\ref{input}). In Fig.\ref{power}(a) we have chosen $\kappa=-1$ (equivalent to $\Omega=1$). Fig.\ref{power}(b) shows the case with $\kappa=-2$, which corresponds to a frequency modulation located at the peak of MI gain (i.e. $\Omega=\sqrt{2}$, the initial FWM phase-matching condition). As can be seen by comparing these two cases, the FPU period $\Lambda$ (and thus the distance of maximal pump depletion obtained at $z=\Lambda/2$) is smaller when the modulation is located at the MI gain peak. On the other hand, a maximum amount of energy is transferred from the pump into the sidebands only when the input modulation frequency is shifted away from the MI gain peak. Indeed, due to the intrinsic pump depletion of the process, phase matching is progressively shifted along the fiber towards lower frequency values, so that the initial modulation with $\Omega=1$ provides the largest pump depletion \cite{capp91,trillo91}.      

We are interested here in studying the MI associated with the spatially periodic steady-state solutions of Eq.(\ref{twe}). We may write a perturbed CW solution of Eq.(\ref{twe}) as
\begin{equation}
\nonumber  A_j(z,t)=\left(\left|\bar{A}_j(z)\right|+a_j(z,t)\right)\exp\{i\phi_j(z)\},
\end{equation}
\noindent where $\left|a_j\right|\ll\left|\bar{A}_j\right|$. By expressing the perturbations $a_j$ as the sum of Stokes and anti-Stokes waves
\begin{equation}
\nonumber  a_j(z,t)=u_j(z)\exp\{i\omega t\}+w_j^*(z)\exp\{-i\omega t\},\label{side}
\end{equation}
\noindent one obtains from Eq.(\ref{twe}) a set of six linear ordinary differential equations with periodic coefficients for the three Stokes and anti-Stokes sideband pairs of the form 
\begin{equation}
\nonumber  \frac{d{\bf X}(z)}{dz}={\bf M}(z){\bf X}(z),\label{odes}
\end{equation}
\noindent where ${\bf X(z)}=(u_0,w_0,u_{-1},w_{-1},u_{1},w_{1})$. In Eq.(\ref{odes}), ${\bf M}(z)$ is a $z$-periodic $6\times6$ matrix with spatial period $z=\Lambda$ equal to the FPU recurrence period. 
By solving Eq.(\ref{odes}) with each of the six fundamental or independent initial conditions: ${\bf X}_1(z=0)=(1,0,0,0,0,0), \, 
{\bf X}_2(z=0)=(0,1,0,0,0,0), \, \ldots, \, {\bf X}_6(z=0)=(0,0,0,0,0,1)$, and by evaluating the corresponding six solutions of Eq.(\ref{odes}) at $z=\Lambda$, one may build the so-called principal solution matrix $S\equiv \{{\bf X}_1^t(z=\Lambda),{\bf X}_2^t(z=\Lambda),\ldots ,{\bf X}_6^t(z=\Lambda)\}$, where the superscript $t$ denotes vector transpose \cite{coddi}. 

According to Floquet theorem, whenever the sideband frequency detuning $\omega$ is such that an eigenvalue of $S$ (or Floquet multiplier) $\lambda_F =\exp(\rho_F+i\sigma)$ is such that $\left|\lambda_F\right|>1$ (or $\rho_F>0$), an instability occurs for the periodic stationary solution of Eq.(\ref{twe}). Namely, the exponential growth with distance of sidebands which are the components of the corresponding eigenvalue ${\bf X}_F$ results. Indeed, since the principal solution or scattering matrix associated with an integer number of periods, say, $z=n\Lambda$ is simply equal to $S^n$, one may relate the Floquet multipliers to the usual MI power gain, say, $\hat{G}$ through the relation $\hat{G}=2\rho_F/\Lambda$. 

The eigenvectors ${\bf X}_F$ of the scattering matrix $S$ associated with unstable eigenvalues $\lambda_F$ are, in general, strongly varying with frequency detuning $\omega$. For any $\omega$, we normalize the eigenvectors of $S$ to unit length (i.e., we set $|{\bf X}_F|^2=1$). For calculating the MI gain associated with each individual sideband, we project $\hat{G}(\omega)$ on the corresponding component of ${\bf X}_F$. The resulting gain for the $k-th$ sideband (with $k=1,2,\ldots,6$) is thus calculated as $G=2\hat{G}|{\bf X}_{F,k}|^2$. Here the factor of $2$ is introduced in order to retrieve the analytical MI gain in the limit case of a constant amplitude CW pump. 

\begin{figure}[ht]
\centering
\includegraphics[width=8.5cm]{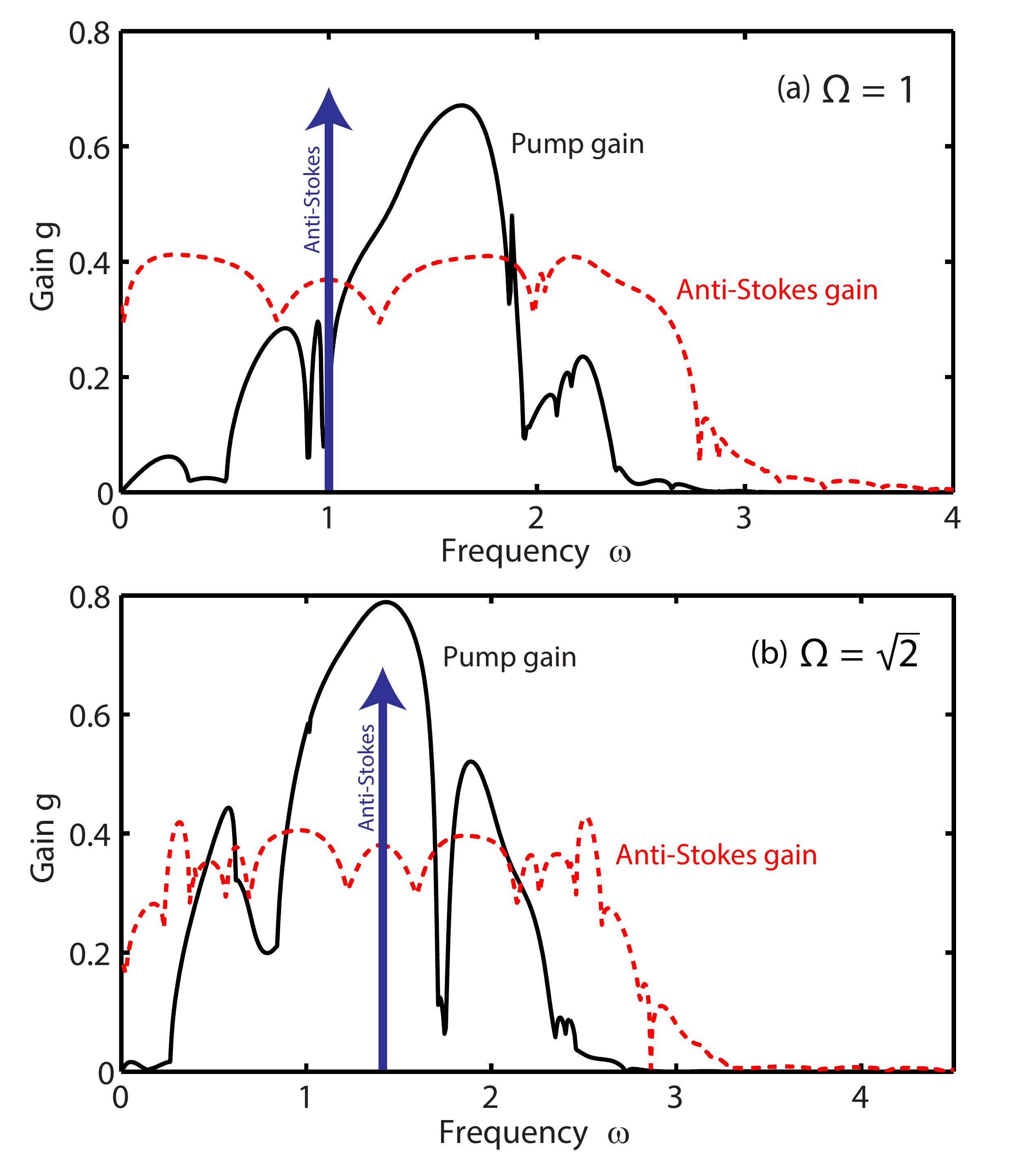} 
\caption{MI gain $g=G/G_m$ for the pump (solid black curves) and anti-Stokes sideband mode (dashed red curves) for: (a) $\Omega=1$; (b) $\Omega=\sqrt{2}$.}
\label{cmi}
\end{figure}

These results are presented in Fig.\ref{cmi}, which illustrates the analytically computed MI gain for either the pump mode $A_0$ (solid black curves) or the anti-Stokes sideband mode $A_{+1}$ (dashed red curves) as a function of the angular frequency shift $\omega$ from the pump, for two different values of the initial modulation frequency $\Omega$ (which is shown in Fig.\ref{cmi} by a vertical blue arrow). Here we present the relative gain $g=G/G_m$, where $G_m=2$ is the peak value of the MI gain for a CW pump (in real units, the peak MI gain is $2\gamma P$, where $\gamma$ is the nonlinear coefficient and $P$ the input CW pump power).

As can be seen in Fig.\ref{cmi}(a) when $\Omega=1$, the MI gain $g$ of the spatially periodic pump mode has a maximum $g_{m \, (\omega = 1.64)} \simeq 0.67$ at $\omega\simeq 1.64$, which is significantly up-shifted compared with the MI gain peak obtained at $\omega=\sqrt{2}$ for a constant amplitude pump. The corresponding gain value, i.e. $g_{m \, (\omega = \sqrt{2})}\simeq 0.6$, is also reduced by about $40\%$ with respect to the constant amplitude pump case. On the other hand, the sideband mode exhibits a gain with a double maxima structure around the sideband frequency associated with a gain peak value $g_{m \, (\omega = 0.25)}\simeq 0.41$. As shown in Fig.\ref{cmi}(b), similar symmetric structures (and gain values) are found for the sideband mode gain when $\Omega=\sqrt{2}$. In this latter case, the pump mode has a slightly larger gain peak value $g_{m}\simeq 0.8$ which is exactly located at $\omega=\sqrt{2}$ (i.e., the MI gain peak frequency obtained for a constant amplitude pump case). In the next section, we compare the results obtained from the linearized stability analysis of the periodic solutions of the truncated Eqs.(\ref{twe}) with the full numerical solutions of the NLSE (\ref{nls1})-(\ref{input}). 

\section{Numerical results}
\label{sec:resu}

In the following, we numerically solved the NLSE (\ref{nls1})  with the initial conditions (\ref{input}), where $\epsilon=0.05$ and $\phi^0=0$, and further added a broadband quantum noise floor (about 125 dB below the spectral power of the background pump) corresponding to the usual one photon per frequency bin with random initial phase in the spectral domain. In our simulations, we implemented a split-step Fourier method with sufficient spatio-temporal discretization to avoid the classical drawbacks associated with unwanted numerical noise generated along propagation. For instance, we used $2^{14}$ time grid points and periodic boundary conditions in time (with a time window equal to $64\, T_\Omega$, where $T_\Omega=2\pi/\Omega$ is the input modulation period) as well as a fixed spatial integration step $\delta z\simeq 1\times10^{-4}$. 

\begin{figure}[ht]
\centering
\includegraphics[width=8.5cm]{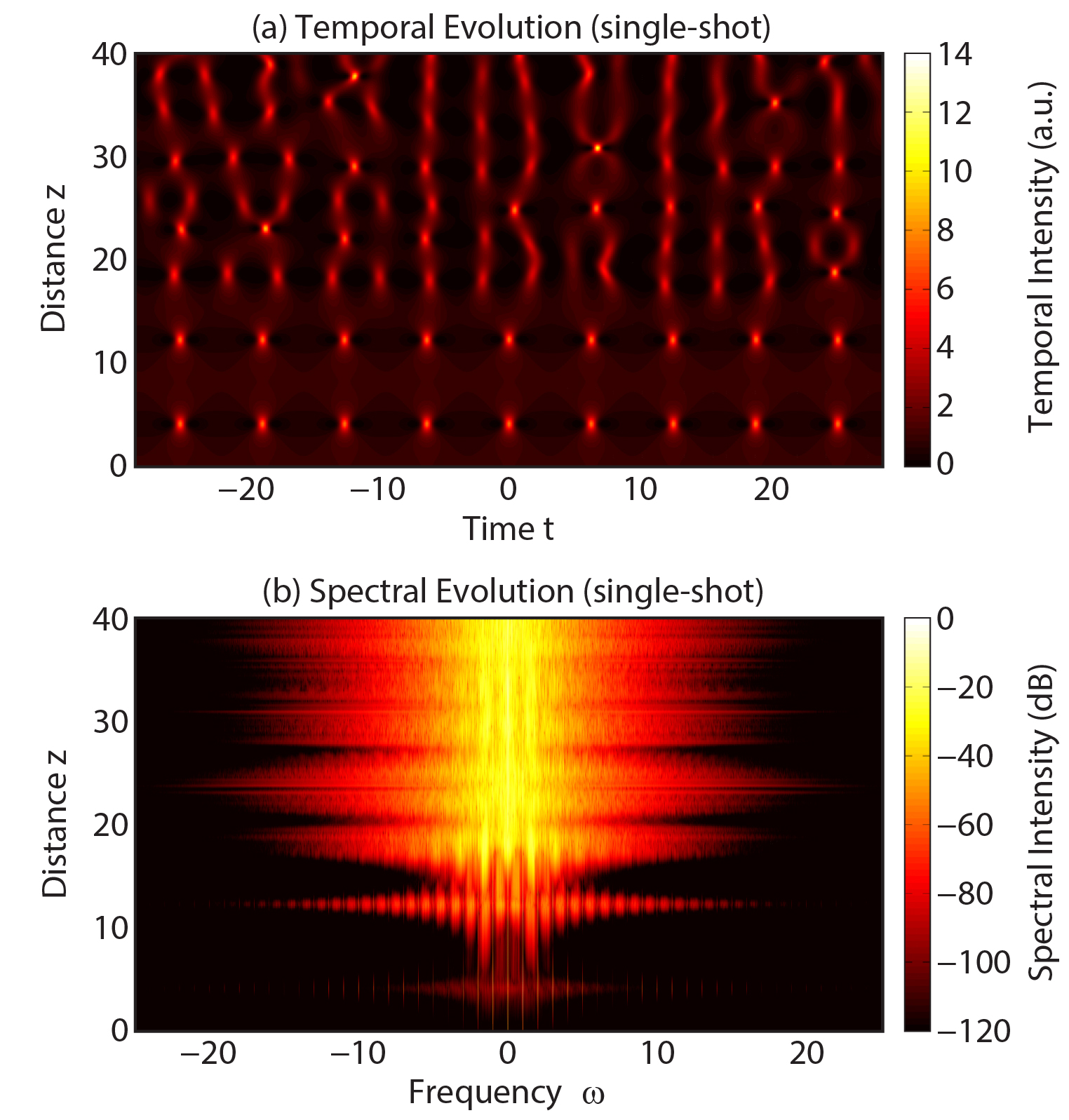} 
\caption{Evolution of single-shot (a) field intensity $|u|^2$ and (b) spectrum (in log scale) for $\Omega=1$.}
\label{amplitudess}
\end{figure}

The contour plot shown in Fig.\ref{amplitudess}(a)  illustrates the spatio-temporal evolution of the field intensity $|u(z,t)|^2$ for the nonlinear phase-matched case $\Omega=1$. Here we show the evolution corresponding to a particular realization of the random input noise seed (single shot case). As it can be seen, after just two FPU recurrence periods, spontaneous MI leads to the field break-up into an irregular structure exhibiting frequency doubling and multiple irregular temporal collapses associated with high intensity peaks formation occurring at different points in space and time. In order to more clearly display the evolution of the individual space-time localized structures, we only show in Fig.\ref{amplitudess}(a) the central temporal profile around $t=0$, as it is extracted from the computed evolution of the $64$ temporal periods of the input modulation. 

Break-up of the FPU recurrence is in fact due to the exponential growth of the initial quantum noise background, as it is induced by the MI of the periodically evolving pump and multiple FWM sidebands. This  is clearly shown in the single-shot spectral domain contour plot of Fig.\ref{amplitudess}(b): here we show the evolution with distance $z$ of the log-scale spectral intensity of the field as a function of the angular frequency detuning from the pump, $\omega$. In correspondence with each temporal compression stage of the FPU process, quantum noise-activated broad sidebands emerge in-between the pump and the first-order seed sidebands, as well as all harmonic (or cascaded FWM) sidebands of the initial modulation at frequencies $\pm n\Omega$, with $n=2,3,4,\ldots$. Fig.\ref{amplitudess}(b) also shows that, after two periods of the FPU recurrence, the temporal field break-up is associated with the growth of a broad frequency continuum among all FWM sidebands, which leads to the irreversible equipartition of energy in frequency space. In correspondence with the formation of high intensity temporal events subsequent to the FPU break-up as shown in Fig.\ref{amplitudess}(a), the associated spectrum exhibits series of sudden frequency broadening (see Fig.\ref{amplitudess}(b)). These spectral features, which are seen at various propagation distances in Fig.\ref{amplitudess}(b) (e.g. $z=24, 31,38$), exhibit an irregular spatial distribution which originates from the amplification of the individual input quantum noise realization.

\begin{figure}[ht]
\centering
\includegraphics[width=8.5cm]{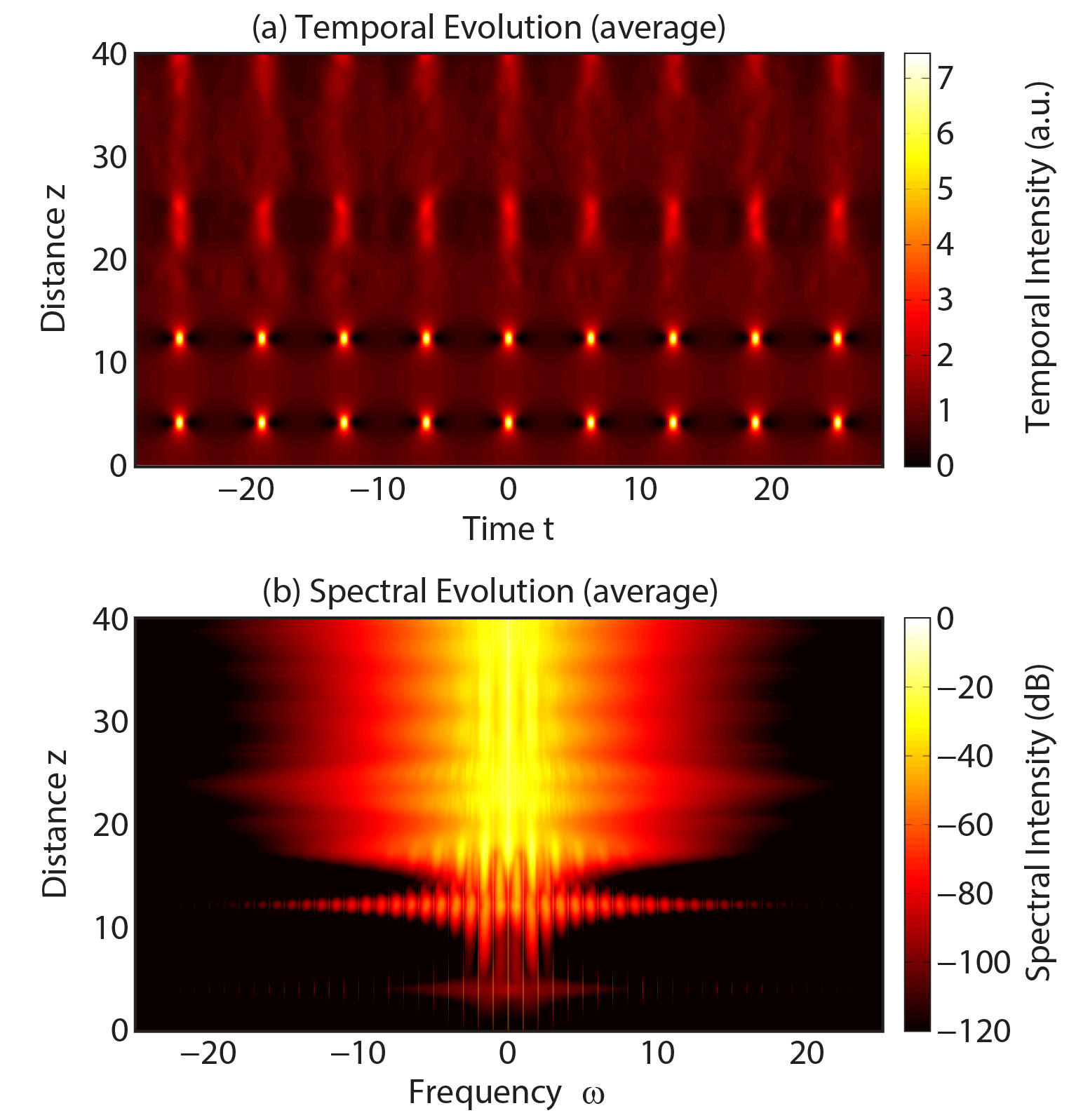} 
\caption{Evolution of noise-averaged (a) field intensity $|u|^2$ and (b) spectrum (in log scale) using 100 realizations for $\Omega=1$.}
\label{amplitudeave}
\end{figure}

In order to gain further insight in the noise growth and associated dynamics in the field propagation, we have also computed noise-averaged evolutions of the field intensity, in both the temporal and spectral domain, as can be seen in Fig.\ref{amplitudeave}. In this case, using the same parameters as those used in Fig.\ref{amplitudess}, averaging was performed over one hundred realizations with independent input quantum noise seeds. As can be seen by comparing Fig.\ref{amplitudess}(a) with Fig.\ref{amplitudeave}(a), the noise-averaged temporal evolution before break-up (i.e. $z<16.5$) is consistent with the single-shot case and the temporal compressions occurring during the two first FPU recurrences do not exhibit significant differences (see temporal intensity scales in both spatio-temporal evolution contour plots). The fact that the main temporal features seen before break-up are not drastically affected by the input noise is further attested by the similar spectral shapes seen up to this point in Fig.\ref{amplitudess}(b) and Fig.\ref{amplitudeave}(b).  

On the other hand, the  averaged spectral evolution after FPU break-up exhibits smooth quasi-periodic breathing compared to the stochastic-like behavior occurring in the single-shot evolution. Such a phenomenon can be understood as follows. The effective input noise amplification of each frequency bin is different for each noise realization. Indeed, the random phase implemented for each input spectral noise component changes the phase matching conditions of the spontaneous MI process. Thus slightly different gain values result for the continuum of noise-seeded frequency components. As a result, the spectrum which is obtained after one (or multiple) FPU recurrence(s) is slightly different for each noise realization. This will therefore impact the complex frequency mixing dynamics which result after FPU break-up. Interestingly, when averaging the spatial evolution of the field after FPU break-up, the sudden spectral broadenings (associated with compressions in the temporal domain) exhibit privileged distances of occurrence (i.e., $z=23,40$), as can be seen in Fig.\ref{amplitudeave}. 


\begin{figure}[ht]
\centering
\includegraphics[width=8.5cm]{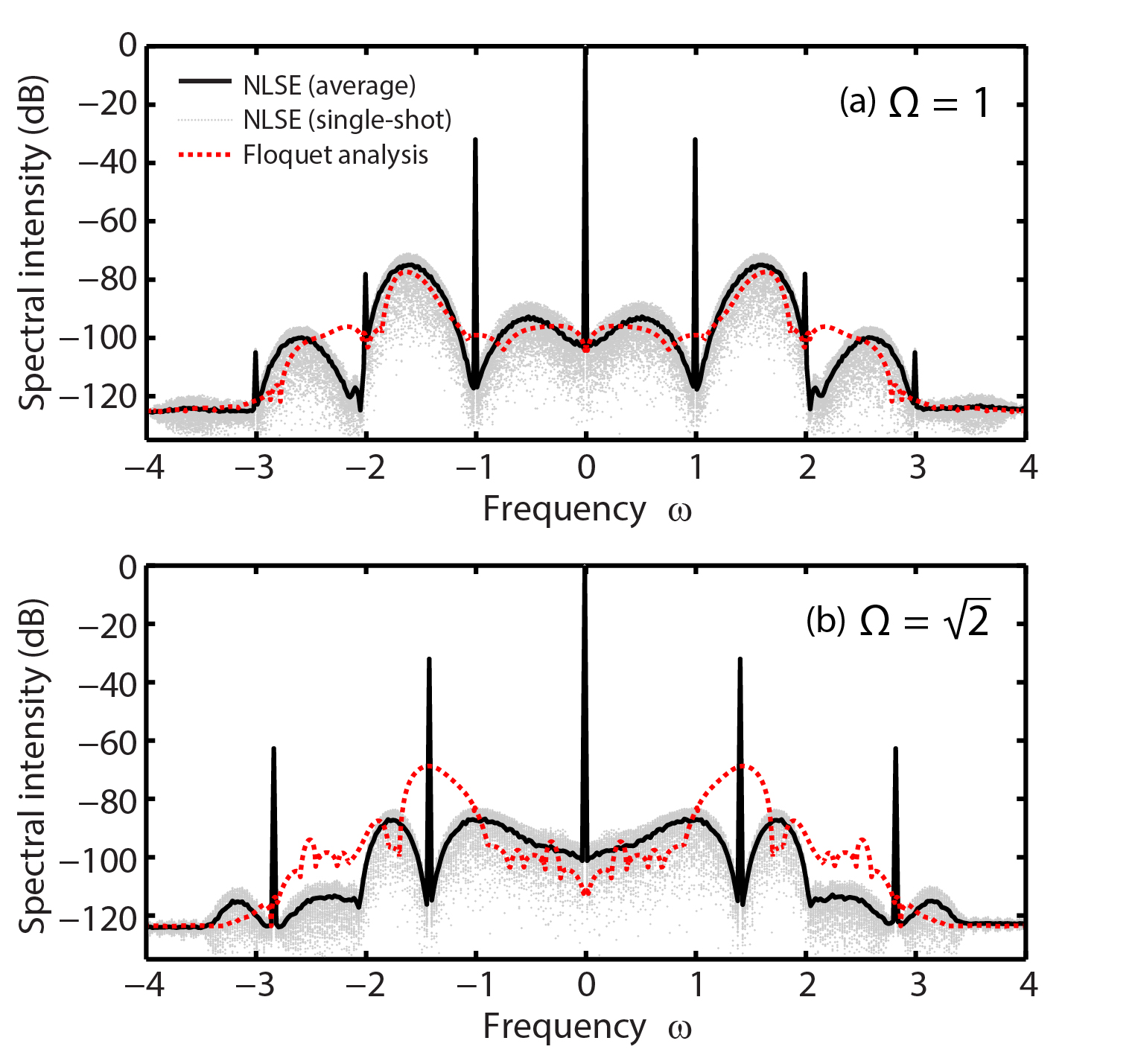}  
\caption{Comparison of spectra from numerical solution of the NLSE (continuous black curves) and from Floquet analysis (dashed red curves) for $z=\Lambda$ and $\Omega=1$ (top); $\Omega=\sqrt{2}$ (bottom). Grey dots indicate the superposition of all 100 single-shot simulated spectra.}
\label{numerical}
\end{figure}

In Fig.\ref{numerical}, we show the spectra which are extracted from the numerical solution of the NLSE (\ref{nls1}) in the nonlinear phase-matched case with $\Omega=1$ (top panel), or in the MI gain peak case $\Omega=\sqrt{2}$ (bottom panel). In both cases, we show the superimposed single-shot spectra obtained from 100 realizations (grey dots), as well as the averaged spectrum (solid black line). All spectra are computed at exactly one spatial period of the FPU recurrence (i.e. $z=\Lambda=8.24$ for $\Omega=1$ and $z=\Lambda=8.07$ for $\Omega=\sqrt{2}$). 

Fig.\ref{numerical} shows that the single-shot spectra exhibit large spectral intensity fluctuations between various realizations, highlighting the impact of the individual initial noise seed on the subsequent amplification process. On the other hand, the average spectra are represented by a smooth black solid curve. The average spectra can be readily compared with the results of the linear stability analysis obtained by the Floquet method presented in section \ref{sec:theo}. The red dashed curves in Fig.\ref{numerical} show the Floquet spectra, which are obtained by considering, for each frequency $\omega$, the maximum between the pump gain and the Stokes or Anti-Stokes gain of Fig.\ref{cmi}.

As can be seen, for $\Omega=1$ there is a good quantitative correspondence between the MI gain spectra obtained from the full NLSE and the Floquet theory applied to the three-wave truncation. Although one may notice slight discrepancies in the surroundings of the initial modulation sidebands and their harmonics, Fig.\ref{numerical}(a) shows that there is an overall agreement between the numerical solutions of the NLSE and the Floquet theory. On the other hand, for $\Omega=\sqrt{2}$ the noise growth predicted by the Floquet analysis near the spectral position of the immediate sidebands is largely overestimated with respect to the NLSE results. In fact the Floquet analysis is not expected to be accurate in this case, because of the resonance occurring when the initial deterministic sidebands (\ref{input}) are superimposed with the position where a spontaneous MI gain peak is predicted (i.e., at $\Omega=\sqrt{2}$). This effectively depletes the spontaneous gain available for quantum noise, thus explaining the lack of agreement in Fig.\ref{numerical}(b) between the numerical solution of the NLSE and the Floquet theory spectra.

\begin{figure}[ht]
\centering
\includegraphics[width=8.5cm]{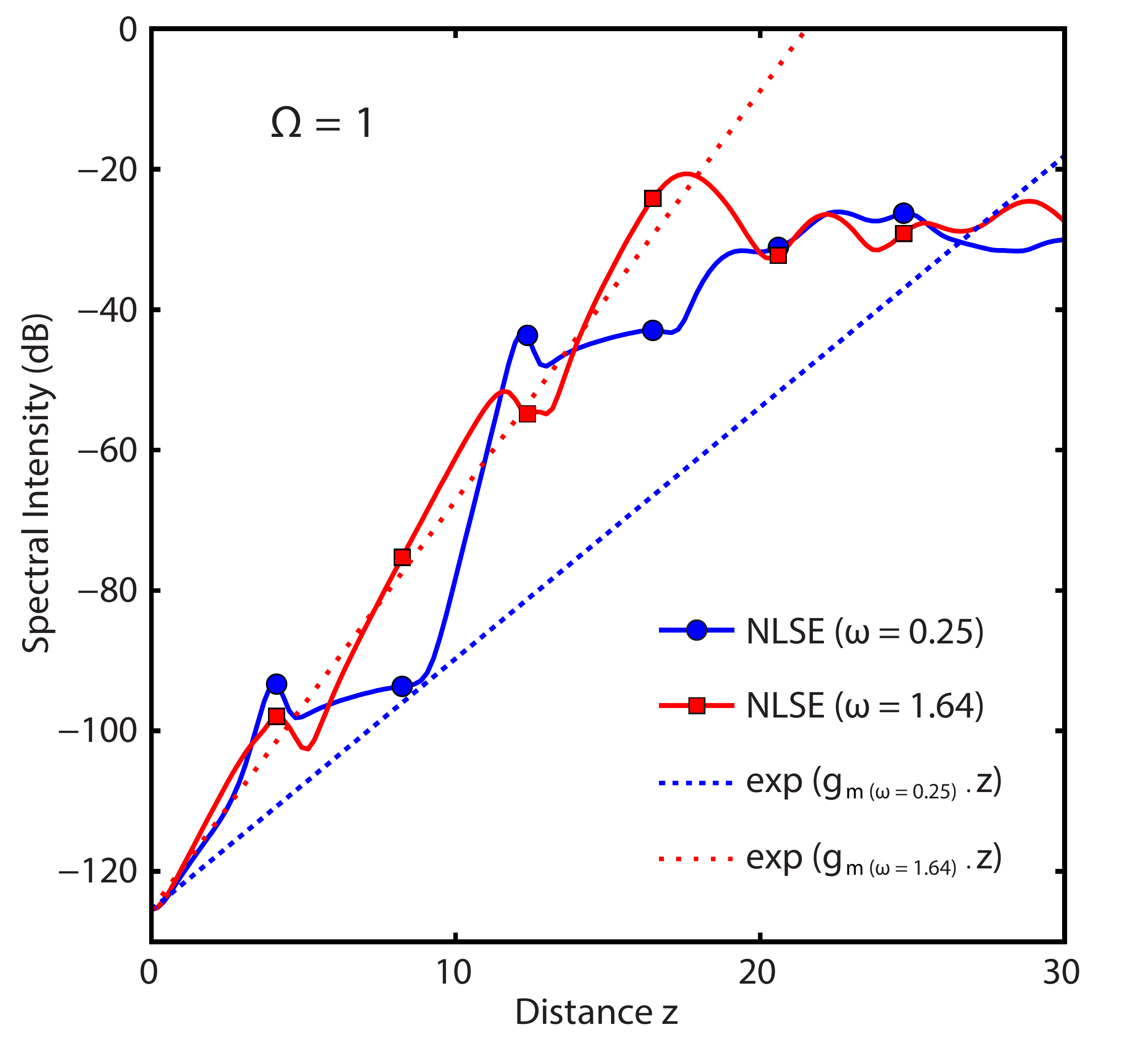} 
\caption{Comparison of Floquet analysis prediction and numerical NLSE results for spectral noise growth with distance for $\Omega=1$. Log scale spectral power is normalized with respect to the input pump power. Dots indicate multiples of the FPU recurrence half-period $\Lambda/2$}
\label{noise}
\end{figure}

Let us compare now the analytical predictions of section \ref{sec:theo} with the spatial growth of spectral noise components which is obtained from the numerical solution of the NLSE (\ref{nls1}) over up to three FPU recurrence periods $z=3\Lambda$. In Fig.\ref{noise} we consider the nonlinear phase-matched case with $\Omega=1$: here we show with a solid blue curve the growth with distance $z$ of the peak (as determined by Floquet gain profile) noise component in the spectral region within the pump and the first anti-Stokes sideband (i.e., with frequency detuning from the pump $\omega=0.25$). Similarly, the solid red curve shows the growth of the maximum noise component between the first and the second anti-Stokes sidebands, i.e., with a frequency shift from the pump $\omega=1.64$. The spatial oscillations which are observed in Fig.\ref{noise} in the growth of the numerically computed peak noise spectral components are due to the periodic temporal compression stages which lead to pump depletion and corresponding cascade FWM generation (see also panels (b) of Figs.\ref{amplitudess}-\ref{amplitudeave}). After two FPU recurrences ($z\simeq 17$), we observe a quasi-stationary evolution (i.e. saturation) of both spectral component intensities, which corresponds to the thermalization process associated with multi-frequency mixing and broadband continuum generation. It is worth mentioning that, in our case, the FPU break-up  occurs after about two FPU periods. Nevertheless, we observed that this phenomenon can actually take place for shorter and longer propagation distances depending on the initial relative modulation sidebands - pump intensities ($\epsilon$) and on the amount of noise added at the input.

For further comparison, the dashed red and blue straight lines in Fig.\ref{noise} illustrate the MI growth which is predicted according to the Floquet gain shown in Fig.\ref{cmi}. For instance, $g_{m \, (\omega = 0.25)} \simeq 0.41$ and $g_{m \, (\omega = 1.64)} \simeq 0.67$. As it can be seen, the Floquet analysis provides a good quantitative fit of the average spatial growth rate of the noise for $\omega=1.64$ until FPU break-up ($z\simeq 17$). On the other hand, the Floquet gain prediction for $\omega=0.25$ is only accurate for the first FPU period $z=\Lambda$. For longer distances, the numerically computed noise growth rate is considerably larger than the analytical predictions. 

\begin{figure}[ht]
\centering
\includegraphics[width=8.5cm]{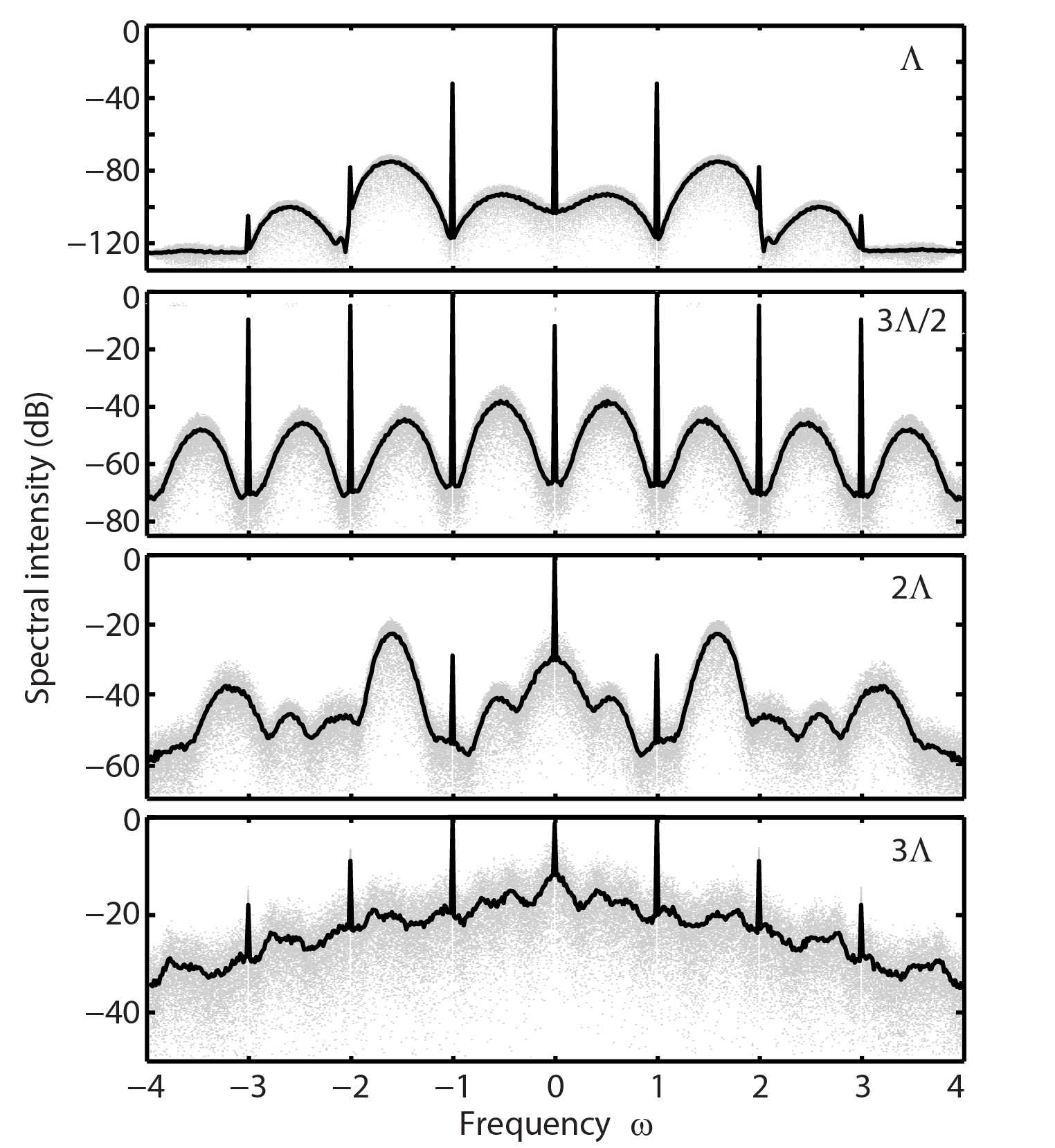} 
\caption{Spectra from the numerical solution of the NLSE for $\Omega=1$ and distances: $z=\Lambda, 3\Lambda/2, 2\Lambda, 3\Lambda$.}
\label{spectra}
\end{figure}

In order to better understand the spectral dynamics of noise growth at distances larger than the first FPU period $\Lambda$,  we have plotted in Fig.\ref{spectra} the output spectra at multiples of the FPU recurrence half period $\Lambda/2$. Quite interestingly, Fig.\ref{spectra} reveals that at the second stage of temporal compression occurring for $z=3\Lambda/2$, the dominating broadband MI noise sidebands centered around $|\omega|\simeq 0.5$ are copied by FWM in multiple mirror images across the entire FWM cascade \cite{arma11,fatome13}. On the other hand, at $z=2\Lambda$ (two FPU recurrence periods ), both the higher order discrete FWM lines and the broadband cascade of MI sidebands disappear, and the spectrum nearly recovers its previous shape seen at $z=\Lambda$. The gain peak enhancement at $|\omega|\simeq 0.5$ (with respect to the gain peak at $|\omega|\simeq 1.5$) observed in at $z=3\Lambda/2$ may be understood as follows. Fig.\ref{spectra} shows that the input modulation sidebands have grown much higher than the original pump mode. Therefore the modulation sidebands at $|\omega|=|\Omega|=1$ temporarily act as pumps, and the corresponding MI gain peaks are approximately obtained at frequency shifts $\pm 1.5$ away from their frequencies. This phenomenon may also explain the analytically underestimated noise growth which is seen in Fig.\ref{noise} for $|\omega| = 0.25$.  Finally, after three FPU recurrence periods ($z=3\Lambda$) a broadband noise continuum has developed: yet, the discrete spectral lines corresponding to the initial frequency comb still remain discernible. 
 
\begin{figure}[ht]
\centering
\includegraphics[width=8.5cm]{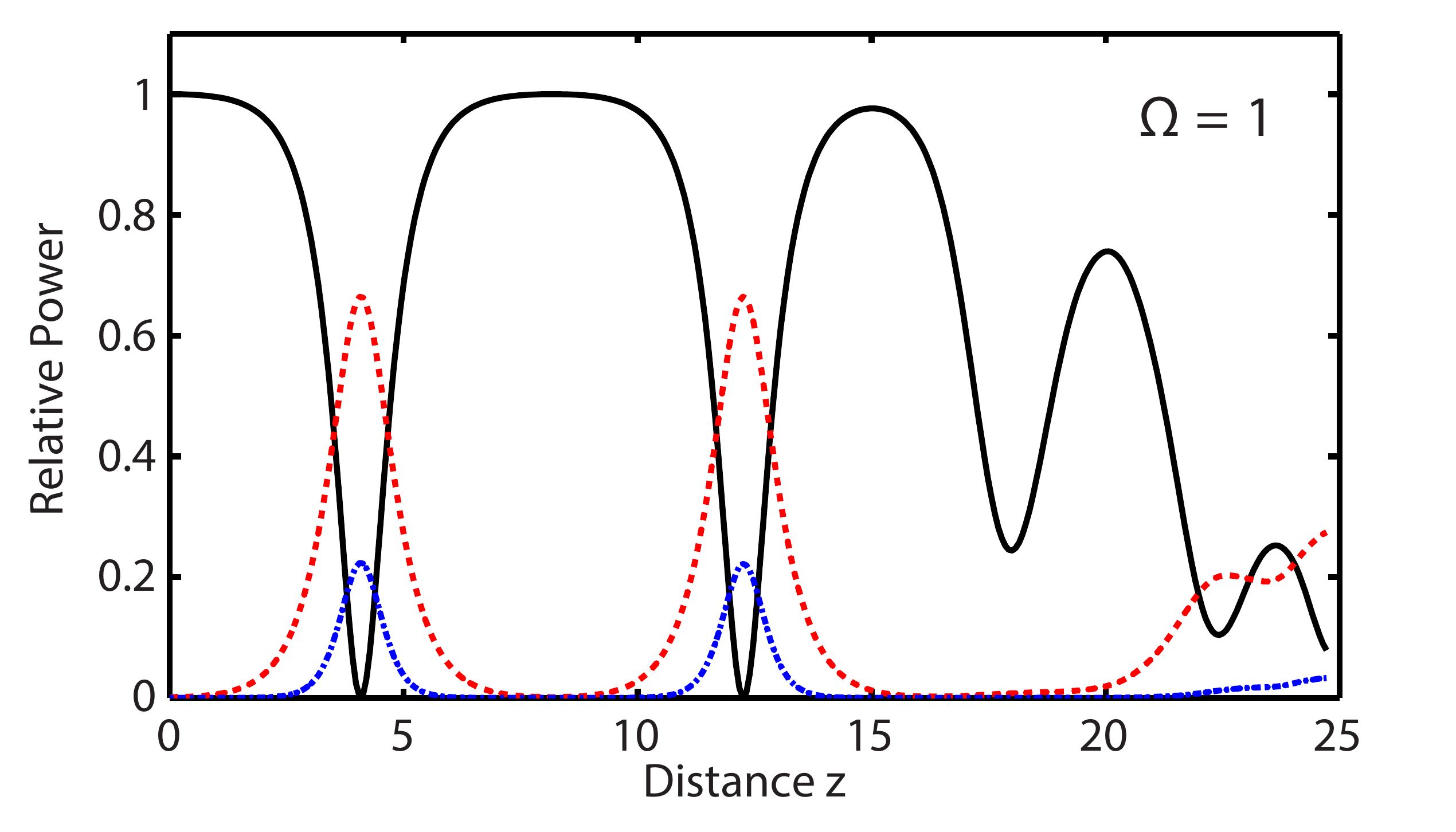}
\caption{Evolution of fractional power of the pump (solid black curve) and the in two sidebands at $\pm\Omega$ (dashed red curve) or $\pm2\Omega$ (dashed blue curve) from the NLSE with $\Omega=1$ plotted over 3 FPU recurrence periods.}
\label{powernum}
\end{figure}

The break-up of the FPU recurrence as it is observed in Figs.\ref{amplitudess}-\ref{amplitudeave} is more clearly displayed in terms of the spatial evolution of the power of the pump and the initial modulation sidebands. Indeed, Fig.\ref{powernum} shows that after two periods of FPU recurrence the pump power suddenly drops down, and it exhibits an irregular evolution around a low average value. In addition, Fig.\ref{powernum} also shows a significant power drop in the immediate sideband powers, as a consequence of the flow of energy from the discrete set of cascade FWM spectral lines into the continuum spectrum. A detailed analysis of the turbulent field behavior resulting after break-up of the FPU recurrence is beyond the scope of this study and will be the subject of further investigations. Nevertheless, it is still interesting to observe that in the context of the purely conservative system presented here, the road through thermalization appears to be a rather complex dynamical process. 

\begin{figure}[ht]
\centering
\includegraphics[width=8.5cm]{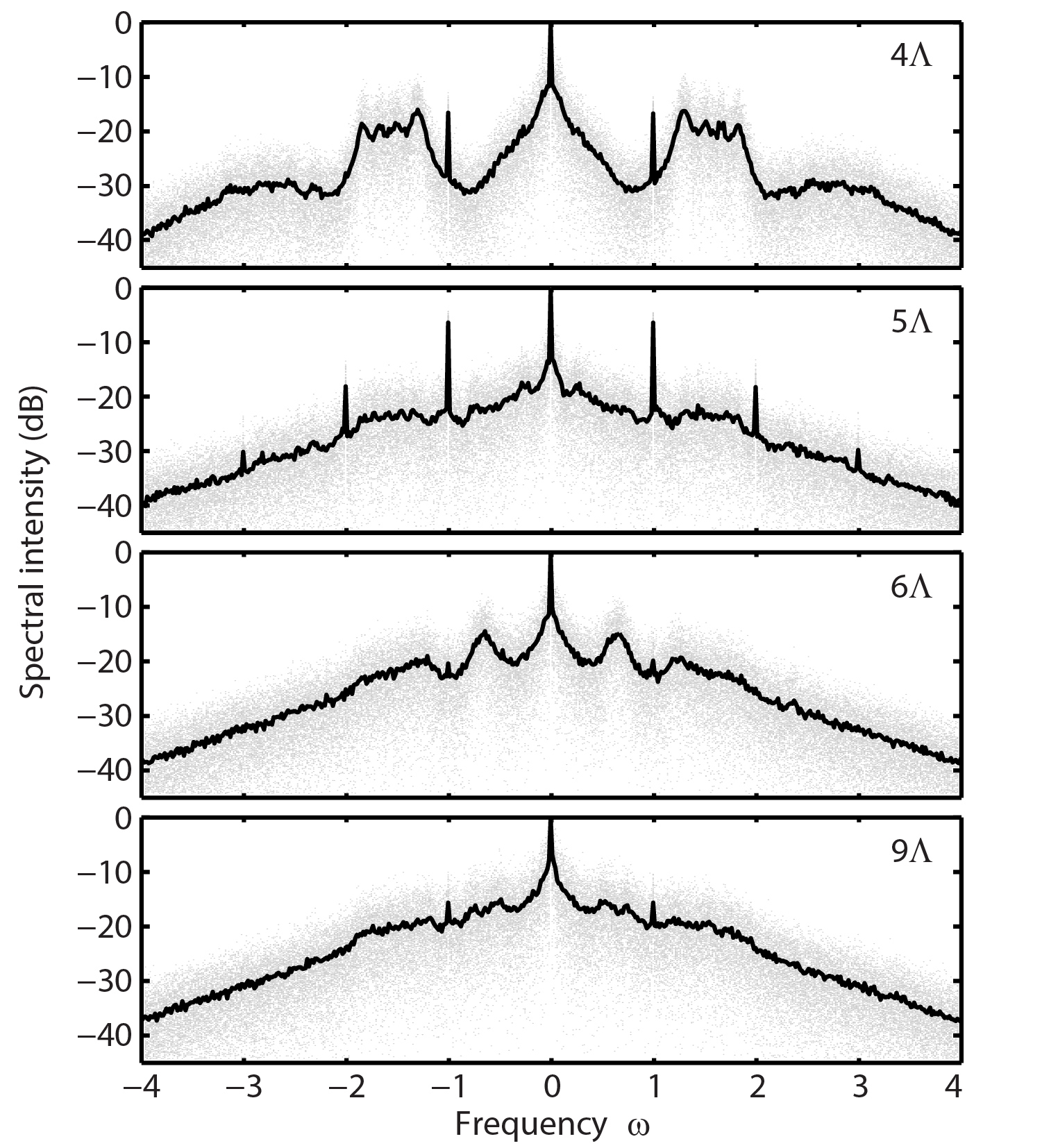}
\caption{Spectra from the numerical solution of the NLSE for $\Omega=1$ and distances: $z=4\Lambda, 5\Lambda, 6\Lambda, 9\Lambda$.}
\label{spectra_therma}
\end{figure}

To illustrate this phenomenon, we present in Fig.\ref{spectra_therma}, in the same way as in Fig.\ref{spectra}, the spectra extracted for longer propagation distances (i.e. $z=4\Lambda, 5\Lambda, 6\Lambda$ and $9\Lambda$). Although we previously observed in Fig.\ref{spectra} the development of a continuum spectrum within the cascaded FWM discrete sidebands at $z=3\Lambda$, spectra obtained for subsequent propagation still exhibit a complex spatial \textit{breathing} associated with a transient enhancement of particular spectral frequencies. Indeed, we can see in Fig.\ref{spectra_therma} at $z=4\Lambda$ the development of broad noise-induced sidebands in the the frequency range $1\leq\left|\omega\right|\leq2 $. At $z=5\Lambda$, the obtained spectrum recovers the shape of a broadband continuum within an FWM cascade. At $z=6\Lambda$ one can see the enhancement of noise sidebands located around $|\omega|\simeq 0.5$. Finally, for even further propagation distances (e.g., $z=9\Lambda$), the \textit{breathing} dynamics start to vanish within the continuum spectrum, which eventually evolves towards quasi-complete thermalization.

\section{Conclusions}
\label{sec:concl}
We presented a theoretical study of instability of the FPU recurrent exchange of power among the pump and its sidebands as described by the self-focusing NLSE. This instability leads to the input noise-induced irreversible evolution of the pump field towards a broadband supercontinuum. As a perspective for further studies, we envisage that the instability of FPU recurrence may also be observed in the presence of relatively small sources of dissipation such as for example a periodically amplified transmission line or a nonlinear dispersive ring cavity, where FPU recurrence break-up may be associated with the development of spatio-temporal turbulence in the frequency comb dynamics.    

\section{Acknowledgments}
Stimulating discussions with C. Finot, B. Kibler, G. Millot, J. M. Dudley and S. Trillo are gratefully acknowledged. The present research was supported by Fondazione Cariplo, grant n.2011-0395, and the Italian Ministry of University and Research (MIUR) (grant contract 2012BFNWZ2).\\






\begin{thebibliography}{99}



\bibitem{bespa} V.I. Bespalov, V.I. Talanov, Pis'ma Zh. Eksp. Teor. Fiz. 3 (1966) 471 (JETP Lett. 3
(1966) 307).

\bibitem{benjamin} T. B. Benjamin, J. E. Feir, J. Fluid Mech. 27 (1967) 417.

\bibitem{dudley06} J. M. Dudley, G. Genty, S. Coen, Rev. Mod. Phys. 78 (2006) 1135.

\bibitem{dudley09} J. M. Dudley, G. Genty, F. Dias, B. Kibler, N. Akhmediev, Optics Express 17 (2009) 21497.

\bibitem{solli07} D. R. Solli, C. Ropers, P. Koonath, B. Jalali, Nature 450 (2007) 1054.


\bibitem{tai} K. Tai, A. Tomita, J.L. Jewell, and A. Hasegawa, Appl. Phys. Lett.  49 (1986) 236.

\bibitem{fermi} E. Fermi, J. Pasta, and S. Ulam, Studies of Nonlinear Problems (Los Alamos Scientific Laboratory Report No. LA-1940, Los Alamos, New Mexico, 1955).

\bibitem{zk} N. J. Zabusky and M. D. Kruskal, Phys. Rev. Lett. 15 (1965) 240.

\bibitem{akhmediev86} N. Akhmediev, V.I. Korneev, Theor. Math. Phys. (USSR) 69 (1987) 1089. Translated from Theor. Mat. Fiz. 69 (1986) 189.

\bibitem{akhmediev88a} N. Akhmediev, V. I. Korneev, and N. V. Mitskevich, Sov. Phys. JETP 67 (1988) 89. Translated from Zh. Eksp. Teor. Fiz. 94 (1988) 159.

\bibitem{akhmediev88b} N. Akhmediev, V. M. Eleonskii and N. E. Kulagin, Theor. Math. Phys. (USSR) 72 (1988) 809. Translated from Teor. Mat. Fiz. 72 (1987) 183.

\bibitem{ablo} M.J. Ablowitz and B.M. Herbst, SIAM J. Appl. Math. 50 (1990) 339.

\bibitem{lake} B. M. Lake, H. C. Yuen, H. Rungaldier, and W. E. Ferguson, J. Fluid Mech. 83, 49 (1977).

\bibitem{chab11} A. Chabchoub, N. P. Hoffmann, and N. Akhmediev, Phys. Rev. Lett. 106 (2011) 204502.

\bibitem{sima01} G.V. Simaeys, Ph. Emplit, and M. Haelterman, Phys. Rev. Lett. 87 (2001) 033902.

\bibitem{sima02} G.V. Simaeys, Ph. Emplit, and M. Haelterman, J. Opt. Soc. Am. B 19 (2002) 477.

\bibitem{tang} B. Zhao, D. Y. Tang, and H. Y. Tam, Optics Express 11 (2003) 3304.

\bibitem{hamma} K. Hammani, B. Wetzel, B. Kibler, J. Fatome, C. Finot, G. Millot, N. Akhmediev, and J.M. Dudley, Optics Lett. 36 (2011) 2140.

\bibitem{kibler10} B. Kibler,	 J. Fatome,	 C. Finot,	 G. Millot,	 F. Dias,	 G. Genty, N. Akhmediev J. M. Dudley, Nature Physics 6 (2010) 790.

\bibitem{kibler12} B. Kibler, J. Fatome, C. Finot, G. Millot, G. Genty, B. Wetzel, N. Akhmediev, F. Dias, J. M. Dudley, Sci. Rep. 2 (2012) 463.

\bibitem{beeckman} J. Beeckman, X. Hutsebaut, M. Haelterman, and K. Neyts, Optics Express 15 (2007) 11185

\bibitem{wu} M. Wu, and C.E. Patton, Phys. Rev. Lett. 98 (2007) 047202.

\bibitem{farota} A.K. Farota and M.M Faye, Phys. Scr. 88 (2013) 055802

\bibitem{infeld81} E. Infeld, Phys. Rev. Lett. 47 (1981) 717.

\bibitem{capp91} G. Cappellini and S. Trillo, J. Opt. Soc. Am. B 8 (1991) 824.

\bibitem{trillo91} S. Trillo and S. Wabnitz, Opt. Lett. 16 (1991) 986.

\bibitem{soto12} J.M. Soto-Crespo, A. Ankiewicz, N. Devine, and N. Akhmediev, J. Opt. Soc. Am. B 29 (2012) 1930.

\bibitem{musso14} A. Mussot, A. Kudlinski, M. Droques, P. Szriftgiser, and Nail Akhmediev, Phys. Rev. X 4 (2014) 011054.

\bibitem{solli12} D. R. Solli, G. Herink, B. Jalali, C. Ropers, Nature Photon. 6 (2012) 463.

\bibitem{wetz12} B. Wetzel,	 A. Stefani, L. Larger,	 P. A. Lacourt,	 J. M. Merolla,	 T. Sylvestre, A. Kudlinski, A. Mussot,	 G. Genty, F. Dias, J. M. Dudley, Sci. Rep. 2 (2012) 882.

\bibitem{dud08} J.M. Dudley, G. Genty, and B.J. Eggleton, Optics Express 16 (2008) 3644.

\bibitem{solli08} D.R. Solli, C. Ropers, and B. Jalali, Phys. Rev. Lett. 101 (2008) 233902.

\bibitem{nguyen13} D. M. Nguyen, T. Godin, S. Toenger, Y. Combes, B. Wetzel, T. Sylvestre, J. M. Merolla, L. Larger, G. Genty, F. Dias, J. M. Dudley, Opt. Lett. 38 (2013) 5338.

\bibitem{trillo97a} S. Trillo and S. Wabnitz, Phys. Rev. E. 56 (1997) 1048.

\bibitem{trillo97b} S. Trillo and S. Wabnitz, Phys. Rev. E. 55 (1997) R4897.

\bibitem{fuerst97} R. A. Fuerst, D. M. Baboiu, B. Lawrence, W. E. Torruellas, G. I. Stegeman, S. Trillo, and S. Wabnitz, Phys. Rev. Lett. 78 (1997) 2756.

\bibitem{arma11} A. Armaroli and S. Trillo, Opt. Lett. 36 (2011) 1999.

\bibitem{fatome13} J. Fatome, C. Finot, A. Armaroli, and S. Trillo, Opt. Lett. 38 (2013) 181.

\bibitem{arma14} A. Armaroli and S. Trillo, J. Opt. Soc. Am. B 31 (2014) 551.

\bibitem{swab10} S. Wabnitz and N. Akhmediev, Opt. Commun. 283 (2010) 1152.

\bibitem{akhmediev85} N. Akhmediev, V.M. Eleonskii, N.E. Kulagin, Sov. Phys. JETP 62 (1985) 894.

\bibitem{erki11} M. Erkintalo, K. Hammani, B. Kibler, C. Finot, N. Akhmediev, J. M. Dudley, and G. Genty, Phys.
Rev. Lett. 107 (2011) 253901.



\bibitem{coddi} E. A. Coddington and N. Levison, Theory of Ordinary Differential Equations (McGraw-Hill, New York, 1955) 78.








 \end{thebibliography}






%
%

\end{document}